%% file: main.tex
\documentclass[lettersize,journal]{IEEEtran}

\ifCLASSINFOpdf
\else
\fi

\usepackage{url}

\usepackage{cite}
\usepackage{hyperref}
\usepackage{amsmath,amssymb,amsfonts}
\usepackage{algorithmic}
\usepackage{graphicx}
\usepackage{textcomp}
\usepackage{comment}
\usepackage{booktabs}
\usepackage[nolist]{acronym}
\usepackage[nolist]{acronym}
\usepackage[normalem]{ulem}
\usepackage{comment,subfigure}
\usepackage{color,soul}
\usepackage{multirow}
\usepackage{multicol,multirow,xcolor,colortbl}
\usepackage{soul}

\newcommand{\bazzi}[1]{\textcolor{black}{#1}}

\newcommand{\festag}[1]{\textcolor{black}{#1}}

\newcommand{\jonas}[1]{\textcolor{black}{#1}}

\newcommand{\miguel}[1]{\textcolor{black}{#1}}

\newcommand{\rev}[1]{\textcolor{black}{#1}}

\begin{document}

\pagenumbering{gobble}
%
\title{Multi-Channel Operation for the Release 2 of ETSI Cooperative Intelligent Transport Systems}


\author{\IEEEauthorblockN{Alessandro Bazzi\IEEEauthorrefmark{1}, 
Miguel Sepulcre\IEEEauthorrefmark{2}, 
Quentin Delooz\IEEEauthorrefmark{3},
Andreas Festag\IEEEauthorrefmark{3},\\ 
Jonas Vogt\IEEEauthorrefmark{4},
Horst Wieker\IEEEauthorrefmark{4}, 
Friedbert Berens\IEEEauthorrefmark{5}, 
 and Paul Spaanderman\IEEEauthorrefmark{6}}

\IEEEauthorblockA{\IEEEauthorrefmark{1}\textit{WiLab, CNIT / DEI, University of Bologna}, Bologna, Italy\\}
\IEEEauthorblockA{\IEEEauthorrefmark{2}\textit{Universidad Miguel Hernandez de Elche (UMH)}, Spain\\}
\IEEEauthorblockA{\IEEEauthorrefmark{3}\textit{Technische Hochschule Ingolstadt, CARISSMA Institute for Electric, COnnected, and Secure Mobility (C-ECOS)}, Germany\\}
\IEEEauthorblockA{\IEEEauthorrefmark{4}\textit{Hochschule für Technik und Wirtschaft des Saarlandes (htw saar), ITS Research Group (FGVT), Germany}\\}
\IEEEauthorblockA{\IEEEauthorrefmark{5}\textit{FBConsulting S.A.R.L}, Luxembourg\\}
\IEEEauthorblockA{\IEEEauthorrefmark{6}\textit{InnoMo}, Monaco}
}


\markboth{}
{Bazzi \MakeLowercase{\textit{et al.}}: Multi Channel Operation for the\\Release 2 of V2X in Europe}
%




\input{Acronyms.tex}

\IEEEtitleabstractindextext{%

\input{Sections/0_Abstract}

}

\maketitle

\IEEEdisplaynontitleabstractindextext

%
\IEEEpeerreviewmaketitle

\acresetall

\input{Sections/1_Introduction}

\input{Sections/2_State-of-the-art}

\input{Sections/3_MCO_Challenges_Principles}
\input{Sections/4_MCO_Concept}

\input{Sections/5_Implementation_options_examples}
\input{Sections/6_Open_issues}
\input{Sections/7_Conclusion}

\section*{Acknowledgment} 
This work has been conducted during the activity of the ETSI Specialist Task Force 585, co-funded by the European Commission.



\bibliographystyle{IEEEtran}  
\bibliography{biblio-papers,biblio-standards}

\end{document}

%% file: acronyms.tex
\begin{acronym} 
\acro{3GPP}{Third Generation Partnership Project}
\acro{4G}{fourth generation}
\acro{5G}{fifth-generation}
\acro{6G}{sixth-generation}
\acro{5GAA}{5G Automotive Association}
\acro{AF}{amplify and forward}
\acro{ALI}{access layer instance}
\acro{AoI}{age of information}
\acro{AS}{application server}
\acro{AWGN}{additive white Gaussian noise}
\acro{BME}{bandwidth management entity}
\acro{B-CSA}{broadcast coded-slotted ALOHA}
\acro{BSM}{basic safety message}
\acro{BS}{base station}
\acro{C-ITS}{cooperative intelligent transport systems}
\acro{C-ITS-S}{cooperative-intelligent transport systems station}
\acro{C-NOMA}{cooperative NOMA}
\acro{C-V2X}{cellular-V2X}
\acro{CAD}{cooperative automated driving}
\acro{CAM}{cooperative awareness message}
\acro{CAS}{cooperative awareness service}
\acro{CBR}{channel busy ratio}
\acro{CCDF}{complementary cumulative distribution function}
\acro{CCH}{control channel}
\acro{CDF}{cumulative distribution function}
\acro{CD}{collision detection}
\acro{CoMP}{coordinated multi-point}
\acro{CP-OFDM}{cyclic prefix orthogonal frequency-division multiplexing}
\acro{CPM}{collective perception message}
\acro{CPS}{collective perception service}
\acro{CRC}{cyclic redundancy check}
\acro{CRDSA}{contention resolution diversity slotted ALOHA}
\acro{CSA}{coded-slotted ALOHA}
\acro{CSI}{channel state information}
\acro{CSIT}{channel state information at the transmitter}
\acro{CSIR}{channel state information at the receiver}
\acro{CSMA/CA}{carrier sense multiple access with collision avoidance}
\acro{D2D}{device-to-device}
\acro{DCI}{downlink control information}
\acro{DEN}{decentralized environmental notification service}
\acro{DENM}{decentralized environmental notification message}
\acro{DMRS}{demodulation reference signal}
\acro{ECP}{extended cyclic prefix}
\acro{EED}{end-to-end delay}
\acro{eNodeB}{evolved NodeB}
\acro{ETSI}{European Telecommunications Standards Institute}
\acro{FCL}{functional configuration limit}
\acro{FCP}{functional configuration profile}
\acro{FD}{in-band full-duplex}
\acro{GAGH}{GeoNetworking ALI group handler}
\acro{GNSS}{global navigation satellite system}
\acro{HARQ}{hybrid automatic repeat request}
\acro{HD}{half-duplex}
\acro{IDMA}{interleave-division multiple access}
\acro{IM}{index modulation}
\acro{ISI}{inter-symbol interference}
\acro{ITS}{intelligent transport system}
\acro{IVI}{in-vehicle-information and other infrastructure-to-vehicle
messages}
\acro{JD}{joint decoding}
\acro{KPI}{key performance indicator}  
\acro{LDPC}{low-density parity-check}
\acro{LOS}{line-of-sight}
\acro{LoA}{Level of Automation}
\acro{LTE}{long term evolution}  
\acro{MAC}{medium access control}
\acro{MAPEM}{MAP (topology) Extended Message}
\acro{MCE}{message collecting entity}
\acro{MCO}{multi-channel operation}
\acro{MCM}{maneuver coordination message}
\acro{MCS}{modulation and coding scheme}
\acro{MGE}{message generating entity}
\acro{MHE}{message handling entity}
\acro{MIMO}{multiple input multiple output}
\acro{MRC}{maximum ratio combining}
\acro{MRE}{message receiving entity}
\acro{MUD}{multiuser detection}
\acro{NLOS}{non-line-of-sight}
\acro{NOMA}{non-orthogonal multiple access}
\acro{NR}{new radio}
\acro{OBU}{on-board unit}
\acro{OFDM}{orthogonal frequency-division multiplexing}
\acro{OFDMA}{orthogonal frequency-division multiple access}
\acro{OMA}{orthogonal multiple access}
\acro{PAM}{platooning awareness message}
\acro{PAS}{position augmentation service}
\acro{PCM}{platooning control message}
\acro{PDB}{packet delay budget}
\acro{PER}{packet error rate}
\acro{PIR}{packet inter-reception}
\acro{PHY}{physical}
\acro{PRB}{physical resource block}
\acro{PRR}{packet reception ratio}
\acro{QAM}{quadrature amplitude modulation}
\acro{QoS}{quality of service}
\acro{RB}{resource block}
\acro{RF}{radio frequency}
\acro{RRI}{resource reservation interval}
\acro{RSRP}{reference signal received power}
\acro{RSU}{roadside unit} 
\acro{SAM}{service announcement message}
\acro{SB-SPS}{sensing-based semi-persistent scheduling}
\acro{SCH}{service channel}
\acro{SCI}{sidelink control information}
\acro{SCMA}{sparse-code multiple access}
\acro{SCS}{subcarrier spacing}
\acro{SDO}{standard development organization}
\acro{SI}{self-interference}
\acro{SIC}{successive interference cancellation}
\acro{SINR}{signal-to-interference-plus-noise ratio}
\acro{SPATEM}{signal phase and timing extended message}
\acro{SR}{scheduling request}
\acro{STF}{Specialist Task Force}
\acro{TB}{transport block}
\acro{TDMA}{time-division multiple access}
\acro{TTI}{transmission time interval}
\acro{UE}{user equipment}
\acro{URLLC}{ultra-reliable and ultra-low latency communications}
\acro{V2N}{vehicle-to-network}
\acro{V2I}{vehicle-to-infrastructure} 
\acro{V2P}{vehicle-to-pedestrian}
\acro{V2V}{vehicle-to-vehicle} 
\acro{V2X}{vehicle-to-everything} 
\acro{VAM}{vulnerable road user awareness message}
\acro{VBS}{VRU basic service}
\acro{VRU}{vulnerable road user}
\acro{VUE}{vehicular user equipment}
\acro{WAVE}{wireless access in vehicular environment}
\acro{WSMP}{WAVE short message protocol}
\end{acronym}

%% file: Sections/0_Abstract.tex
\begin{abstract} 
Vehicles and road infrastructure are starting to be equipped with \ac{V2X} communication solutions to increase road safety and provide new services to drivers and passengers. In Europe, the deployment is based on a set of Release~1 standards developed by ETSI to support basic use cases for \ac{C-ITS}\acused{ITS}. For them, the capacity of a single 10~MHz channel in the \ac{ITS} band at 5.9~GHz is considered sufficient. At the same time, the \ac{ITS} stakeholders are working towards several advanced use cases, which imply a significant increment of data traffic and the need for multiple channels. 
\miguel{To address this issue, ETSI has recently standardized a new \ac{MCO} concept for flexible, efficient, and future-proof use of multiple channels. This new concept is defined in a set of new specifications that represent the foundation for the future releases of \ac{C-ITS} standards. The present paper provides a comprehensive review of the new set of specifications, describing the main entities extending the \ac{C-ITS} architecture at the different layers of the protocol stack, In addition, the paper  provides representative examples that describe how these MCO standards will be used in the future and  discusses some of the main open issues arising.} The review and analysis of this paper facilitate the understanding and motivation of the new set of Release~2 ETSI specifications for MCO and the identification of new research opportunities.
\end{abstract}

%% file: Sections/1_Introduction.tex
\section{Introduction} 
\label{sec:introduction}

Cooperative, connected and automated mobility (CCAM) will require the use of wireless communications to contribute to the ``Vision Zero'' of the EU, which targets no road deaths by 2050. In the past years, various organizations including IEEE, ETSI, SAE, ISO, and 3GPP have developed different standards to enable direct data exchange among vehicles, other road users, and the infrastructure. In Europe, the effort has resulted in a set of ETSI specifications implementing the Release~1 of \ac{C-ITS}\acused{ITS} as listed in ETSI~TR~101~607.\footnote{ETSI standards are available free of charge at \url{https://www.etsi.org}.}

Release 1 covers so-called ``Day-1'' applications~\cite{6979970,7911287}, 
based essentially on the exchange of 
 \acp{CAM}, sent repetitively by each vehicle to inform about their status and movements, and \acp{DENM}, sent on an event basis to warn about safety-critical situations. Due to the limited amount of shared data, a single 10~MHz radio channel was regarded as sufficient, and Release~1 standards were not designed to support the simultaneous use of multiple channels. 

\input{TableMessages}

\input{TableStandards}

The emergence of new  applications\footnote{In C-ITS, messages are generated by applications or by entities called services, which are implemented at the facilities layer. To improve readability, in this paper we use the term \textit{application} to include both sources of messages.} that go beyond \mbox{``Day-1''} motivate the creation of the ETSI Release~2 set of \ac{C-ITS} standards. With Release~2, road users will 
share information about the surrounding environment, using collective perception (ETSI~TS~103~324), will 
create platoons of vehicles (ETSI~TR~103~299) or coordinate their maneuvers 
(ETS~ITS~103~561). Vulnerable road users (i.e., 
bicycles, scooters, etc.) will also generate messages to inform about their presence (ETSI~\mbox{TS~103~300-3}). The messages generated and the estimated number of channels needed are summarized in Table~\ref{tab:messages}, from which it is clear that a single channel is not sufficient. 

Release~2 will require several channels, possibly using more than one transceiver and more than one radio access technology  \cite{Spectrum5GAA,SpectrumC2C20}.  
Given the necessity to define rules for the use of multiple channels, \miguel{ETSI has recently approved a set of    specifications about \ac{MCO}}. These standards, presented in Table~\ref{tab:standards}, define how the various entities inside the \ac{C-ITS} station collect information and make decisions to use multiple channels.  
\jonas{To enable efficient management of multiple channels, the new set of specifications adds to the C-ITS station architecture a new core entity acting at the facilities layer. This entity collects information about the implemented applications with their requirements and the available radio access technologies. It is designed to control and negotiate various settings to optimize channel utilization and ensure compliance with application requirements. Additional entities at the networking \& transport and the access layers, and the corresponding internal communication flows, are defined to allow software components to be developed by different stakeholders, thus ensuring a modular implementation of C-ITS stations. This new set of ETSI standards represents one of the first steps towards Release~2, although  certain aspects are left for future specifications, such as} the definition of the channel to be used by each application, or the rules for the amount of traffic that can be allocated to a given channel and offloaded to another.

This paper \miguel{provides a comprehensive review and analysis of the ETSI specifications on MCO and} the deriving MCO concept, describing the main entities of the C-ITS architecture and their operation and interactions. \miguel{To the authors knowledge, this is the first paper that performs this review and analysis.} In addition, representative examples that explain how this standardized solution will be used are discussed, \miguel{which are important for future implementations and the definition of profiles. Finally,} new research opportunities for the future exploitation of the MCO concept are discussed. \bazzi{To help the reader, the main acronyms used in the paper are added in Table~\ref{tab:acronyms}.}

\input{TableAcronyms.tex}

%% file: TableMessages.tex
\begin{table*}
	\caption{Main C-ITS messages expected in Release~2.}
	\label{tab:messages}
		\centering 
		\thispagestyle{empty}
	\begin{tabular}{p{7.1cm}p{5cm}p{3cm}p{1.3cm}}
\toprule
\textbf{Message type} & \textbf{Scope} &  \textbf{\hbox{Traffic characteristics} (*)} &  \textbf{\hbox{\# channels} (**)} \\
\midrule
Cooperative awareness messages (CAMs)\acused{CAM} & Continuous notification of status and movements from vehicles & 1-10 Hz, 400 bytes& 0.9 \\
Decentralized environmental notification messages (DENMs) \acused{DENM} & Notification of specific events & 1-10 Hz, 350-1000 bytes & 0.1 \\
Signal phase and timing messages (SPATs)\acused{SPAT} & Intersections and traffic management from & 10-50 Hz, 1200 bytes & 0.5 \\
~~~ \& MAP (topology) messages (MAPs)\acused{MAP} & road-side units & & \\
Vulnerable road user awareness messages (VAMs)\acused{VAM} & Continuous notification of status and movements from pedestrians, bicycles, scooters, and other vulnerable road users & 1-10 Hz, 350 bytes & 0.5 \\
Platooning control messages (PCMs)\acused{PCM} & Platoon internal management& 50 Hz, 400 bytes& 1 \\
Collective perception messages (CPMs)\acused{CPM} & Sharing of sensor-based perception of the surrounding & 1-10 Hz, 1000 bytes & 2 \\
Maneuver coordination messages (MCMs)\acused{MCM} & Coordination of cooperative manoeuvres & 1-10 Hz, 1000 bytes & 2 \\
\bottomrule
\multicolumn{3}{l}{(*) Estimated rate and size of the messages, as from \cite{SpectrumC2C20}}\\\multicolumn{3}{l}{(**) Estimated number of 10~MHz channels occupied by the given messages, as  from \cite{SpectrumC2C20}}\\
\end{tabular}
\end{table*}

%% file: TableStandards.tex
\begin{table*}
	\caption{Set of ETSI standards on MCO for \ac{C-ITS} Release~2.}
	\label{tab:standards}
		\centering 
		\thispagestyle{empty}
	\begin{tabular}{p{2cm}p{4.0cm}p{10.8cm}}
\toprule
\textbf{ETSI Standard} & \textbf{Scope for MCO} & \textbf{Content} \\
\midrule
TR 103 439 & Technical report on MCO & Report discussing the context, motivations, and possible MCO approaches. Also includes simulation results for adjacent channel interference in the 5.9~GHz band. \\
TS 103 696 & Extension of the \ac{C-ITS} architecture
& Specification extending the C-ITS communication architecture defined in ETSI EN 302 665. \\
TS 103 697 & Architecture & Standard defining the architecture of MCO with the entities and main functionalities.\\
TS 103 141 & Facilities layer part & Standard defining the entities and functionalities related to MCO at the facilities layer.\\
TS 103 836-4-1 & Networking \& transport layer part & GeoNetworking standard that extends the the media-independent functionalities by MCO.\\
TS 103 695 & Access layer part & Standard defining the entities and functionalities related to MCO at the access layer.\\
\bottomrule
\end{tabular}
\end{table*}

%% file: TableAcronyms.tex
\begin{table}
	\caption{\bazzi{Main acronyms and abbreviations.}}
	\label{tab:acronyms}
	\centering 
	\thispagestyle{empty}
	\begin{tabular}{p{1cm}p{6cm}}
    \toprule
    C-ITS & Cooperative intelligent transport system \\
    MCO & Multi-channel operation\\
    \midrule
    \multicolumn{2}{l}{\textbf{Main C-ITS-generic acronyms and abbreviations}} \\
    ACC & Access \\
    CAM & Cooperative awareness message \\ 
    CAS & Cooperative awareness service \\
    CPM & Collective perception message \\
    CPS & Collective perception service \\
    FAC & Facilities \\
    NET & Networking \& transport \\
    SAM & Service announcement message\\
    \midrule
    \multicolumn{2}{l}{\textbf{Main MCO-specific acronyms}} \\
    ALI & Access layer instance \\
    BME & Bandwith management entity \\
    FCP & Functional configuration profile \\
    GAGH & GeoNetworking ALI group handler \\
    MCE & Message collecting entity \\
    MGE & Message generating entity \\
    MHE & Message handling entity \\
    MRE & Message receiving entity \\
    \bottomrule
    \end{tabular}
\end{table}

%% file: Sections/2_State-of-the-art.tex
\section{State of the Art}\label{sec:soa}

\subsection{\festag{Regulation and }\miguel{Standardization} }

\begin{figure*}[t]
\centering
\includegraphics[scale=0.5, draft=false]{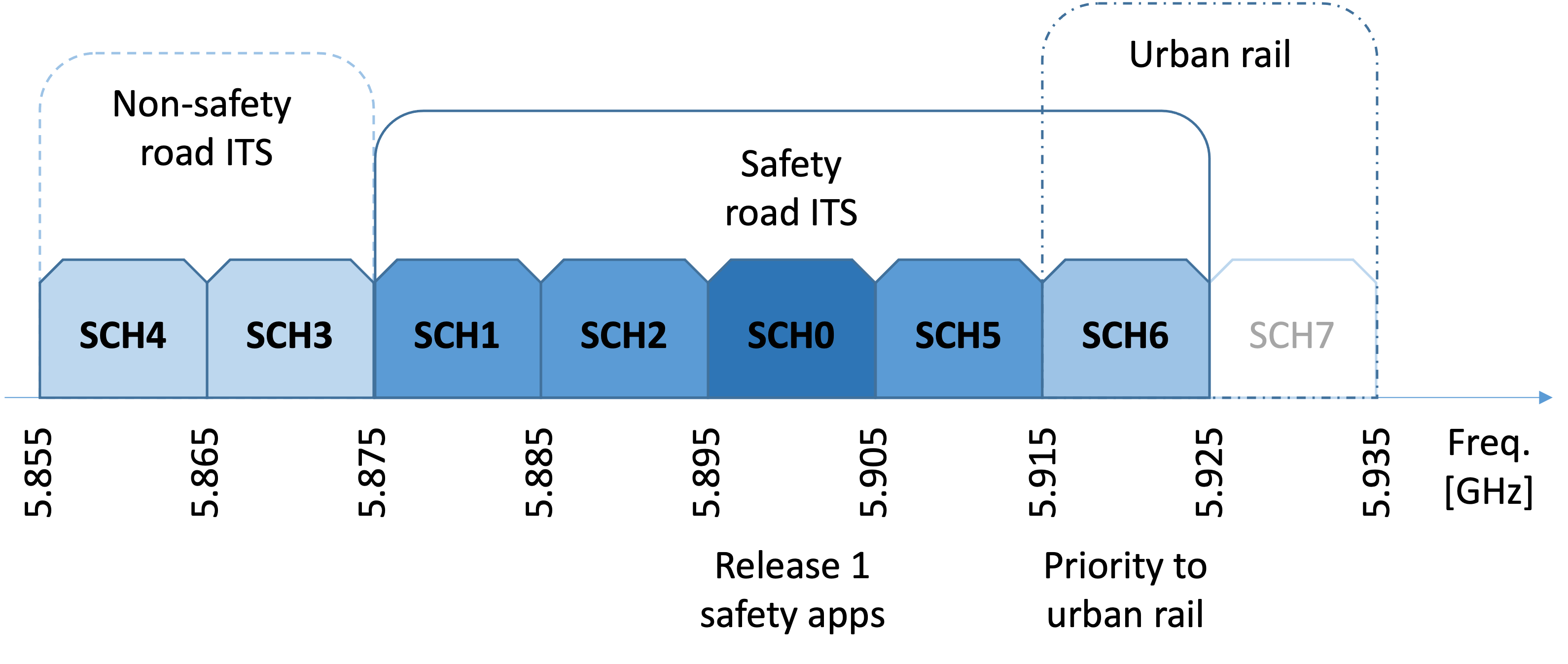}
\caption{Allocation of the channels in the ITS band around 5.9 GHz in Europe.}
\label{fig:channels}
\end{figure*}

\miguel{Different standardization bodies such as ETSI, SAE, ISO, and IEEE have specified architectures, protocols, and services to enable V2X communications. They all rely on} the spectrum reserved for direct \ac{V2X} communications, which mainly consists of a group of 10~MHz channels in the 5.9~GHz band, also known as \ac{ITS} band. The latest regulations\footnote {Implementing Decision 2020/1426 of the European Commission, October 2020.} for this band in Europe reserve for road applications seven 10~MHz channels named \acp{SCH}\bazzi{, as shown in Fig.~\ref{fig:channels}}. One of the channels (SCH0) is the channel primarily used by the Release~1 safety applications (including \acp{CAM} and \acp{DENM}), which was originally referred to as \ac{CCH}. Two of the channels are assigned as non-safety channels and one is subject to priority for urban rail systems.

\miguel{The reserved ITS spectrum for direct \ac{V2X} communications is agnostic to radio technologies. Currently, two main families of standards are available~\cite{8723326}. The first family is based on IEEE~802.11p and its enhancement IEEE~802.11bd. It relies on \ac{CSMA/CA} and it is fully distributed and asynchronous. The second family has been specified by the 3GPP and includes LTE-V2X sidelink and NR-V2X sidelink.   
These 3GPP technologies make use of multi-carrier-based multiple access and the radio resource allocation can be distributed or controlled by the network, with the former case normally considered for the ITS band. On the C-V2X roadmap, the next steps are the integration of pedestrians and enhanced QoS.} To date, in Europe, no restrictions have been defined for the use of the channels by either technology. Work has been done to investigate the co-channel coexistence and mitigate the mutual interference (ETSI TR~103~667 and ETSI TR~103~766); however, an agreement has not yet been reached.

Prior to ETSI, different standards for MCO have been published. IEEE specified an MCO scheme in IEEE~1609.4 as part of the \ac{WAVE} standards \cite{Kenney_DSRCstandards}. WAVE includes channel coordination, channel routing, and QoS parameter mapping using the enhanced distributed channel access (EDCA) in the data plane. The management plane specifies several MCO-related services, among which 
multi-channel synchronization of WAVE devices via Time Advertisement frames and 
channel access control. 
IEEE~1609.4 relies on two types of channels, i.e. the \ac{CCH} and the \acp{SCH}. The CCH may be used by the \ac{WSMP} only. SCHs may be used by \ac{WSMP} and/or IPv6. 
In contrast to the MCO framework \miguel{defined by ETSI and reviewed in this paper}, IEEE~1609.4 is specific to a single access technology and tightly linked to the WAVE standards.

\miguel{Also, ISO defined in ISO~17423 the parameters and processes required to support automatic selection of communication profiles. A communication profile is a parameterized protocol stack, including parameters such as the channel and the flow type. The complete set of parameters includes operational, destination, communication performance, security and protocol communication service parameters. The ISO standards have partially inspired the ETSI specifications on MCO, but do not integrate mechanisms to handle channel overloads.}

\miguel{ETSI specifications for the Release~1 of C-ITS assumed the use of a single radio channel and radio access technology. The emergence of new C-ITS applications and services in Release~2 impose the need of multiple channels and radio access technologies, and thus the definition of MCO specifications. }

\subsection{Research}


Besides standardization, MCO has been considered in research. In~\cite{Haerri2015} and \cite{Campolo_MCO_servey}, the authors provide an overview of the MCO-related standardization activities at that time. 
They also describe the channel allocation and switching principles, including synchronous and asynchronous approaches, based on distributed channel management and \acp{SAM}. \jonas{In \cite{9079451}, based on \ac{WAVE}, the focus is on the use of two transceivers with one tuned to the CCH and the other one to the SCH that optimizes the multi-hop routing of messages. 
The authors in \cite{7875465}, assuming a single transceiver again with \ac{WAVE}, define a flexible scheduling algorithm for safety and non-safety messages controlled by the \acp{RSU}. 
In \cite{s19153283}, multiple channels are considered for the design of a new MAC protocol 
that selects a non-shared or a shared channel for each transmission, with the scope to maximize the 
throughput. 
All three proposals do not take into account the application requirements and have either a single transceiver or two transceivers with one fixed to the CCH.} 
The authors of \cite{BOBAN201617} present an approach for service-actuated multi-channel operation for vehicular communications (SAMCO) with asynchronous channel switching. 
They use \acp{SAM} to distribute service information, whereas a station can prioritize the selection of services based on user preferences and channel load. SAMCO assumes that all systems are equipped with dual transceivers, with one transceiver continuously listening to the CCH, and assumes that services are in pre-defined categories. The authors in~\cite{Sepulcre_CarHet} describe a technology and application-agnostic, distributed, context-aware heterogeneous \ac{V2X} communication system (CARHet).
The proposed channel selection algorithm allows every station to dynamically choose the appropriate access technology and channel based on the application requirements. 
CARHet assumes that all vehicles can transmit and listen to all channels (i.e. they have as many radio interfaces as channels) and focuses on balancing the load among the different channels to minimize interference and packet losses. All these existing MCO schemes have been considered and taken into account for developing \miguel{the ETSI MCO framework. However, standards must take into account the regulation restrictions, as well as existing deployments and future evolution, which challenge the specification process}.

%% file: Sections/3_MCO_Challenges_Principles.tex
\section{MCO Requirements and Principles}\label{sec:challenges} 

\subsection{Requirements}
Many requirements must be taken into account by MCO for the successful deployment of Release~2 applications. One of the main ones is the \bazzi{\textit{coexistence of C-ITS applications with different characteristics and needs}}. The MCO should consider these needs (e.g., priority, latency, or bandwidth) for efficient and effective spectrum utilization, which is particularly important given the safety nature of most C-ITS applications.

Another key challenge is that the \bazzi{\textit{application needs will change over time}}. For example, the frequency and relevance of some messages increase when a vehicle approaches an intersection. The efficient exploitation of the radio channels requires the dynamic adaptation of MCO and close interaction between the MCO entities and the C-ITS applications.

An aspect also to consider is that \bazzi{\textit{different C-ITS stations may implement different applications and have different capabilities}} (e.g., number of radio interfaces). The MCO concept needs to ensure the efficient and fair coexistence of, e.g., simple C-ITS stations implementing only a few advanced applications in a limited number of channels, and advanced C-ITS stations capable of using all the radio channels and implementing a wide range of applications. 

MCO should also be \bazzi{\textit{backward compatible}} with Release~1 solutions. In particular, it must consider that Release~1 C-ITS stations make use of SCH0 for the transmission and reception of messages and that their generation and transmission are subject to congestion control rules. 

Finally, C-ITS applications \bazzi{\textit{mainly rely on the exchange of broadcast messages}}. While the use of the broadcast transmission mode significantly simplifies certain aspects, it has important implications for the design of MCO. The main one is that the selection of a channel by the transmitting station needs to be known by the intended destinations. This means that some kind of coordination is needed, which may for example require the exchange of control information or an agreed association between channels and applications. 

\subsection{Principles}\label{subsec:principles}
As detailed in ETSI TR~103~439, the MCO concept must define a channel usage mechanism and a channel association policy. The \textit{channel usage mechanisms}, dealing with the order of use of the channels, can be classified into the following three approaches:  
\begin{itemize}
  \item \textbf{Sequential filling:} the channels are used in a predefined order. Consequently, a given channel is not used until the prior channels are fully loaded. 
\item \textbf{Load balancing:} the channel use aims at balancing the load among the existing channels. It is considered for example in CARHet~\cite{Sepulcre_CarHet}. \item \textbf{Elastic:} there is no restriction in the order nor the load distribution over the radio channels. It is used, for example, in SAMCO~\cite{BOBAN201617}.
\end{itemize}

The \textit{channel association policies}, defining how each application associates its messages to a channel for their transmission, can be classified as follows:
\begin{itemize}
\item \textbf{Predefined association policies:}
the channel associated with each application is predefined. It naturally fits with the elastic channel usage since the channel load cannot be easily balanced or ordered when the channels to be used are predefined. 
\item \textbf{Flexible association policies:} each application can individually select the channel for the transmission of its messages. This type of policy could be used with any of the above-described channel usage mechanisms.
\end{itemize}

The first set of MCO standards (see Table~\ref{tab:standards}) is designed to permit all the listed channel usage mechanisms and channel association policies. Which combination to use will be defined in future application specifications or in MCO profiles agreed upon between stakeholders. The elastic channel usage with predefined association presents several advantages compared to the other alternatives: 
one key benefit is predictability because each message type is associated with a certain channel and no additional signaling is required between stations; a second one is that it allows stakeholders to implement the number of radio interfaces that are necessary to run the applications that they want to support, enabling the coexistence of diverse implementations; finally, it enables backward compatibility by simply associating Release~1 messages to SCH0.

%% file: Sections/4_MCO_Concept.tex
\section{The MCO Concept}\label{sec:concept}

\begin{figure*}
    \centering
    ~~~\includegraphics[scale=0.4]{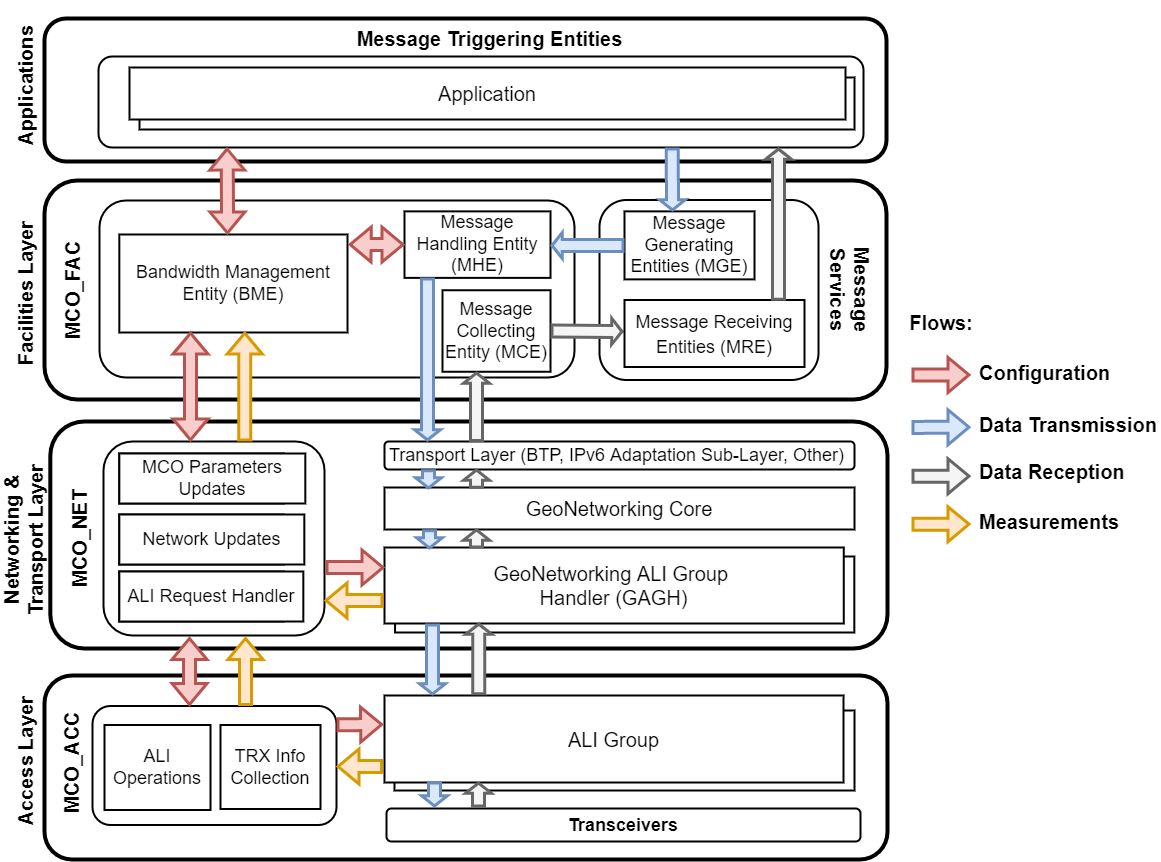}
    \caption{Scheme of the MCO architecture internal of the C-ITS station.}
    \label{fig:architecture}
\end{figure*}

\subsection{Architecture}

The MCO components of the C-ITS architecture are depicted in Fig.~\ref{fig:architecture}. As observable, MCO has components in all the layers: MCO$\_$FAC at the facilities, MCO$\_$NET at the networking \& transport, and MCO$\_$ACC at the access layer. These components are in turn composed of entities dedicated to specific purposes and interact with the other components of the C-ITS station, from the applications 
to the physical transceivers. 

\subsection{The core is at the facilities layer}

A key principle of the MCO framework is that MCO decisions are performed at the facilities layer; therefore, MCO$\_$FAC represents the main part of the concept. This is necessary because the decisions (i) need to consider both the requirements of the applications and the actual status at the access layer, (ii) require a full view of the currently executed applications. The former aspect makes the C-ITS remarkably different from systems such as cellular networks. It considers that the C-ITS station is part of an ad-hoc network and does neither act as a base station nor user equipment. The latter is due to the need to deal with data traffic related to the safety of life, which cannot be treated as best-effort traffic.

\subsection{Functional configuration profiles}

The needs of the applications are communicated to the MCO$\_$FAC through \textit{\acp{FCP}}. The \ac{FCP} sets the requirements, which may include, e.g., the estimated data rate, the maximum latency, and the minimum one-hop range. It is proposed by the application but needs confirmation by MCO$\_$FAC, which knows the actual capabilities at the lower layers and sets \acp{FCL}. The FCP and FCL can eventually be the result of a negotiation between the MCO$\_$FAC and the applications. The FCP of an application can change over time due to different needs of the application or variations at the lower layers. It can also be different for separate data flows of the same application, for example, because part of the messages is safety-critical, and part is supplementary.

\subsection{Access layer instances}

One or more transceivers can be present at the lower layers, potentially based on different technologies. The possible configuration of a transceiver, including the specific access technology type, channel, and \ac{MCS}, is called \textit{\ac{ALI}}. Several \acp{ALI} corresponding to the same access technology and channel are called \textit{\ac{ALI} group}. Each \ac{ALI} in a group can be active or not. 
The \ac{ALI} represents a media-independent abstraction of the transceiver capabilities.  The group concept enables a transceiver to send and receive messages with different configurations; however, each message can only be associated with a single \acs{ALI}. Measurements, such as the channel load, are performed on an \acs{ALI} group basis. 

\subsection{MCO procedures}
\label{subsec:MCOprocedures}

The MCO procedures consist of four groups of operations. The flows internal to the C-ITS station are indicated through arrows in Fig.~\ref{fig:architecture}. 

\textbf{Application resource allocation:} At its initialization, an application requests resources from the \textit{\ac{BME}} within the MCO$\_$FAC. The BME evaluates the available resources and determines those that can be allocated to the requesting application. Besides returning the decision to the application, the BME also informs another entity within MCO$\_$FAC, called the \textit{\ac{MHE}}, which is in charge of internally routing the data from higher to lower layers. 
The BME also commands the settings of the \acp{ALI} and \acs{ALI} groups to MCO$\_$NET, based on the possible configurations at the lower layers and the requirements of the applications. The MCO$\_$NET forwards the settings to MCO$\_$ACC through the \textit{\ac{GAGH}}, which realizes channel-dependent, media-dependent, and media-independent GeoNetworking functionalities. The allocation of application resources or the \acs{ALI} configuration can later be modified following any kind of variations communicated by the higher or lower layers. 

\textbf{Data transmission:} The traffic generated by an application via the \textit{\ac{MGE}} is managed by the MHE within MCO$\_$FAC based on the associated FCPs, available \acp{ALI}, and the settings provided by the BME. The MHE may withdraw or offload messages to other channels whenever needed, in which case it also notifies the BME. 

\textbf{Data reception:} The frames received at the access layer are passed by the receiving ALI up to the corresponding \ac{GAGH}. The GAGH delivers the message to the \textit{\ac{MCE}}, which is in charge of distributing the content to the applications for which it is relevant via the \textit{\ac{MRE}.} 

\textbf{Updates from the lower layers:} Measurements about channel occupation are continuously performed at the access layer and reported by MCO$\_$ACC to MCO$\_$NET, which in turn reports it in a technology-agnostic way to the MCO$\_$FAC. The MCO$\_$NET can also report to MCO$\_$FAC measurements received from the neighboring stations. Additionally, the lower layers can notify unexpected events like the withdrawal of a message from the transmission queue, for example, due to exceeding the maximum latency.

%% file: Sections/5_Implementation_options_examples.tex
\section{C-ITS Station Implementation Examples}
\label{sec:examples}

\begin{figure*}[t]
\centering
\includegraphics[width=0.95\textwidth]{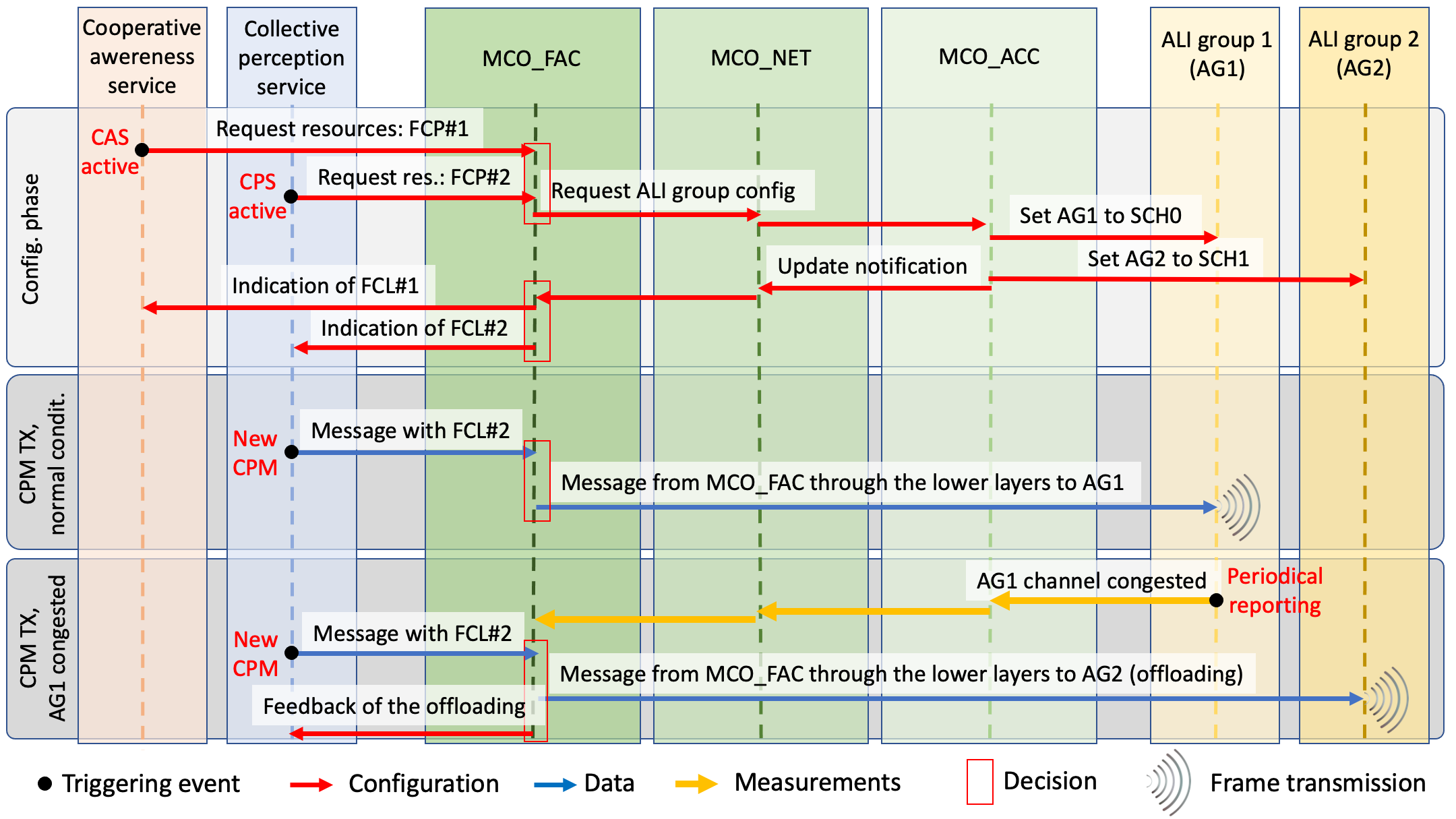}
\caption{\bazzi{Internal communications related to the MCO procedures when the example of option~2 (Sec.~V-B) is considered, which assumes two applications (cooperative awareness and collective perception services) served through two transceivers, corresponding to two ALI groups. Three procedures are exemplified: the configuration phase, during which the FCLs are negotiated and the transceivers tuned to two different channels; the transmission of a message in the primary channel under normal channel conditions; and the offloading of a message on a different channel when a congestion status is measured in the primary channel.}}
\label{fig:example}
\end{figure*}

The flexible MCO framework allows various options for C-ITS station implementation with respect to the number of applications, transceivers, and channels. This section discusses two examples of variants.


\subsection{Single transceiver and data generated by one application}

In this example, the C-ITS station has a single ITS-G5 transceiver and activates the \ac{CAS} as the only application generating messages. For backward compatibility, the transceiver operates in the SCH0. At initialization, based on a preconfigured \ac{FCP}, the MCO$\_$FAC activates all MCO-related components, following the MCO procedures. For operation, CAS generates CAMs and requests their dissemination to the MHE. If sufficient resources are available, MHE passes the messages to the lower layers with the determined ALI. If the resources are insufficient, the MHE informs the \ac{CAS} of the limitation of the resources. Furthermore, the MHE may discard some messages and communicate this event to the CAS.

\subsection{Dual transceiver and data generated by two applications}
\label{subsec:example2}

In this example, the station has two NR-V2X transceivers and activates two applications that generate messages, the \ac{CAS} and the \ac{CPS}. \bazzi{The internal procedures related to this example are detailed in Fig.~\ref{fig:example}.} At initialization, 
the CAS negotiates resources in SCH0, with messages discarded in case of insufficient resources. The CPS requests the activation of an FCP for the transmission of messages with lower priority than CAS, using SCH0 as a preferred channel and SCH1 as an alternative channel in the case of congestion. \bazzi{The BME gives instructions to the lower layers to tune the two ALI groups (i.e., the two transceivers) to SCH0 and SCH1, and returns the FCLs to the two services. During normal operation, both services send their messages through the same ALI group, tuned to SCH0.} In case of congestion \bazzi{(inferred by the periodical reporting performed by each ALI group)}, the CPS offloads some of the messages to \bazzi{the ALI group 2, tuned to} SCH1. This is realized by the MCO$\_$FAC, which passes these messages with a different ALI to the lower layers and provides feedback to the CPS. With growing congestion level, different actions may be taken utilizing the interaction between MCO$\_$FAC and the applications: (i) the CPS can reduce the message rate, (ii) the CPS can intelligently discard messages, and (iii) the CAS can discard some of its lower priority messages.



%% file: Sections/6_Open_issues.tex
\section{\bazzi{Discussion and} Open Issues}
\label{sec:openissues}

\bazzi{The MCO concept detailed in Section~\ref{sec:concept} has its core in the MCO$\_$FAC entity, which collects (and possibly negotiates) the requirements from the applications and configures the lower layers accordingly. This mechanism, which was not present in Release~1, introduces the use of multiple channels but also of multiple technologies and manages conditions where the resources are insufficient to serve all implemented applications.  \jonas{MCO} brings modifications to all layers of the C-ITS protocol stack and implies new issues and challenges in research, standardization, and deployment, among which we highlight the 
following aspects.}

\textbf{Application-driven message prioritization:} In Release~1, the messages generated by an application are assumed with the same priority. In some cases, their generation frequency can change based on simple rules that try to limit the channel occupation. With the described concept, applications can be designed with more complex criteria, associating different \acp{FCP} to the different messages depending on aspects that are known only by the application. This may lead to priority being defined on a per-message basis to better capture the relevance of the carried information.

\textbf{Channel usage and association}: 
The assignment of services to the available channels in the 5.9~GHz frequency band requires coordination across all C-ITS stations\bazzi{, following the principles detailed in Section~\ref{subsec:principles}. 
It may correspond initially to a simple scheme with predefined channel associations and later include advanced solutions that dynamically adapt to the vehicular context and balance the traffic over the channels. The choice of the specific scheme to adopt will require collaboration between all 
stakeholders. Among the challenges to be considered is that, in practical scenarios, differently configured C-ITS stations are expected to coexist, which may have a different number of radio interfaces.} 

\textbf{MCO congestion control}: Because of the safety-related requirements, mechanisms to manage channel congestion should be mainly at the functional level, i.e., in the facilities layer. In Release~1, congestion control is instead only realized by discarding messages at the access layer \cite{Smely2015}.
With more channels, multiple access technologies, and various additional applications, the access layer does not have the required comprehensive information. \bazzi{For Release~2, it is envisioned that the access layer should be used as a strict enforcer of the legislation to access the channel, i.e. the Radio Equipment Directive (RED) and the harmonized ETSI EN~302~663. The MCO$\_$FAC entity will intelligently manage the available radio resources among the applications and will be able to react to congestion in one channel by offloading the traffic to another one or requesting the application to reduce the generated traffic. The definition of optimal decisions depends on the context and will require further work.} 

\bazzi{\textbf{MCO interference}: An issue that needs to be considered jointly with congestion control is the interference between adjacent channels. 
}\miguel{
Given the non-ideality of communication equipment, the transmissions performed in one channel cause unwanted emissions on nearby channels, which impact differently depending on the channel separation. It has been shown, e.g. in~\cite{Campolo_ACI}, that the interference is negligible when channels are separated by a gap of 10~MHz or more. In contrast, communication reliability reduces when the channels are adjacent. It is shown in ETSI~TR~103~439, through simulations assuming ITS-G5 in a highway scenario, that the maximum distance between source and destination to have 90\% packet reception probability reduces by up to 40\% when two adjacent channels are highly loaded.} \bazzi{Limitations to the load in one channel could thus provide a benefit also to the adjacent ones.}

\bazzi{\textbf{Exchange of perceived channel status}: In Release~1, ITS-G5 enables the exchange of the locally perceived channel status with other stations. However, this option was practically not deployed. For Release~2, it is expected that this status exchange will become more relevant. For example, even if a transceiver of a station is not tuned to a channel, other stations could provide information regarding its occupation; this information could then be used to predict future conditions and realize better decisions on which channel to tune its transceivers. This aspect, like the others described in this section, requires further studies.} 

\festag{\textbf{Management overhead and complexity}: Compared to the single channel of Release~1, the MCO concept facilitates the efficient utilization of the whole spectrum, which comes at a cost. In addition to the new MCO-related functions, the MCO concept affects all layers of the protocol stack (see Fig.~\ref{fig:architecture}) and contributes to increased management traffic internally to the C-ITS station. Moreover, it could imply additional information exchange also between stations, for example through the use of \acp{SAM}. 
It is 
worth noting that the presented MCO concept is flexible since it allows for a trade-off between the implemented MCO functionality and complexity, ranging from a single transceiver and single application to several ones (see Sec.~\ref{sec:examples}).}

\rev{\textbf{Implementation issues}: 
As said, the presented MCO framework facilitates the implementation of complex station variants with several transceivers (possibly with different access technologies), accessing several channels, and, more importantly, with a large number of applications. While the concept supports a theoretically infinite number of transceivers and applications, it can be expected that economic and business aspects will limit their number. From a practical perspective, the MCO framework requires the definition of \acp{ALI} and \acs{ALI} groups available for every transceiver and the setting of \acp{FCP} for every application. It is worth noting that these configurations need to be consistent within the station and should realize one of the concepts detailed in Section~\ref{subsec:principles}, with application prioritization and mapping between services, transceivers, and channels, as well as inputs from the research conducted on the other topics discusses in this section.}

%% file: Sections/7_Conclusion.tex
\section{Conclusion}
\label{sec:conclusion}

\bazzi{In this paper, we have reviewed the specifications and 
discussed the new MCO concept introduced in Release~2 of C-ITS \miguel{specifications for the exploitation of multiple radio channels and access technologies}.} \miguel{MCO evolves} the existing C-ITS architecture with functionalities on all layers \miguel{of the protocol stack}. Its key features are: (i) an access layer abstraction that allows for a flexible operation of radio transceivers, (ii) an intelligent MCO functionality at the facility layer that manages the efficient usage of the bandwidth and the generation of messages, and (iii) an effective cross-layer interaction among the functionalities. While the MCO framework builds a cornerstone for the future release of C-ITS standards, it also raises new challenges for research, standardization, and deployment\miguel{, that are discussed in this paper}.